\documentclass[%
 pra,
 reprint,
 superscriptaddress,
 amsmath,amssymb,
 aps
]{revtex4-1}
\usepackage{graphicx}
\usepackage{dcolumn}
\usepackage{bm}
\usepackage{bbm}

\usepackage[utf8]{inputenc}
\usepackage[english]{babel}
\usepackage[T1]{fontenc}
\usepackage{amsmath}
\usepackage{amssymb}
\usepackage{hyperref}
\usepackage{float}
\usepackage{booktabs}
\usepackage{multirow}
\usepackage{graphicx}
\usepackage{enumitem}
\usepackage{listings}
\usepackage{soul}
\usepackage{tikz}
\usepackage{lipsum}
\usepackage{physics}
\usepackage{bm}
\usepackage{array}
\usepackage[mathlines]{lineno}
\usepackage{comment}

\begin{document}

\preprint{APS/123-QED}

\title{Hybrid quantum learning with data re-uploading on a small-scale superconducting quantum simulator}

\author{Aleksei Tolstobrov}
\email{tolstobrov.ae@phystech.edu}
\affiliation{Laboratory of Artificial Quantum Systems, Moscow Institute of Physics and Technology, 141700 Dolgoprudny, Russia}
\affiliation{Russian Quantum Center, 121205 Skolkovo, Moscow, Russia}
\author{Gleb Fedorov}
\affiliation{Laboratory of Artificial Quantum Systems, Moscow Institute of Physics and Technology, 141700 Dolgoprudny, Russia}
\affiliation{Russian Quantum Center, 121205 Skolkovo, Moscow, Russia}
\affiliation{Laboratory of Superconducting Metamaterials, National University of Science and Technology ‘MISIS’, 119049 Moscow, Russia}
\author{Shtefan Sanduleanu}
\affiliation{Laboratory of Artificial Quantum Systems, Moscow Institute of Physics and Technology, 141700 Dolgoprudny, Russia}
\affiliation{Russian Quantum Center, 121205 Skolkovo, Moscow, Russia}
\affiliation{Laboratory of Superconducting Metamaterials, National University of Science and Technology ‘MISIS’, 119049 Moscow, Russia}
\author{Shamil Kadyrmetov}
\affiliation{Laboratory of Artificial Quantum Systems, Moscow Institute of Physics and Technology, 141700 Dolgoprudny, Russia}
\author{Andrei Vasenin}
\affiliation{Center for Engineering Physics, Skolkovo Institute of Science and Technology, 121205 Moscow, Russia}
\affiliation{Laboratory of Artificial Quantum Systems, Moscow Institute of Physics and Technology, 141700 Dolgoprudny, Russia}
\author{Aleksey Bolgar}
\affiliation{Center for Engineering Physics, Skolkovo Institute of Science and Technology, 121205 Moscow, Russia}
\affiliation{Laboratory of Artificial Quantum Systems, Moscow Institute of Physics and Technology, 141700 Dolgoprudny, Russia}
\author{Daria Kalacheva}
\affiliation{Center for Engineering Physics, Skolkovo Institute of Science and Technology, 121205 Moscow, Russia}
\affiliation{Laboratory of Artificial Quantum Systems, Moscow Institute of Physics and Technology, 141700 Dolgoprudny, Russia}
\affiliation{Laboratory of Superconducting Metamaterials, National University of Science and Technology ‘MISIS’, 119049 Moscow, Russia}
\author{Viktor Lubsanov}
\affiliation{Laboratory of Artificial Quantum Systems, Moscow Institute of Physics and Technology, 141700 Dolgoprudny, Russia}
\author{Aleksandr Dorogov}
\affiliation{Laboratory of Artificial Quantum Systems, Moscow Institute of Physics and Technology, 141700 Dolgoprudny, Russia}
\author{Julia Zotova}
\affiliation{Laboratory of Artificial Quantum Systems, Moscow Institute of Physics and Technology, 141700 Dolgoprudny, Russia}
\affiliation{Laboratory of Superconducting Metamaterials, National University of Science and Technology ‘MISIS’, 119049 Moscow, Russia}
\author{Peter Shlykov}
\affiliation{Laboratory of Artificial Quantum Systems, Moscow Institute of Physics and Technology, 141700 Dolgoprudny, Russia}
\author{Aleksei Dmitriev}
\affiliation{Laboratory of Artificial Quantum Systems, Moscow Institute of Physics and Technology, 141700 Dolgoprudny, Russia}
\affiliation{Laboratory of Superconducting Metamaterials, National University of Science and Technology ‘MISIS’, 119049 Moscow, Russia}
\author{Konstantin Tikhonov}
\affiliation{L. D. Landau Institute for Theoretical Physics, 142432 Chernogolovka, Russia}
\author{Oleg V. Astafiev}
\affiliation{Center for Engineering Physics, Skolkovo Institute of Science and Technology, 121205 Moscow, Russia}
\affiliation{Laboratory of Artificial Quantum Systems, Moscow Institute of Physics and Technology, 141700 Dolgoprudny, Russia}

\date{\today}

\begin{abstract}
Supervised quantum learning is an emergent multidisciplinary domain bridging between variational quantum algorithms and classical machine learning. Here, we study experimentally a hybrid classifier model using quantum hardware simulator -- a linear array of four superconducting transmon artificial atoms -- trained to solve multilabel classification and image recognition problems. We train a quantum circuit on simple binary and multi-label tasks, achieving classification accuracy around 95\%, and a hybrid quantum model with data re-uploading with accuracy around 90\% when recognizing handwritten decimal digits. Finally, we analyze the inference time in experimental conditions and compare the performance of the studied quantum model with known classical solutions.
\end{abstract}

\maketitle

\section{Introduction}

Over the last years, high attention has been attracted by the idea of using parameterized quantum circuits (PQC) as universal approximating models for machine learning (ML) tasks \cite{mitarai2018quantum, schuld_1, havlivcek2019supervised, jerbi2023quantum}, inspired by research on variational quantum algorithms (VQA). There are many types of VQA \cite{cerezo2021variational}, from which quantum neural networks (QNN) seem to be the most attractive for solving classical ML problems. While various architectures for QNNs have been suggested, including convolutional \cite{cong2019quantum}, generative-adversarial \cite{dallaire2018quantum, lloyd2018quantum}, recurrent \cite{bausch2020recurrent} networks, it is still not clear how to overcome the general trainability issues  \cite{mcclean2018barren, wang2021noise, anschuetz2022quantum, holmes2022connecting} when such models have high quantum volume; this is an area of active research \cite{volkoff2021large, mele2022avoiding, du2022quantum}. In recent works it has been shown, that a small-scale QNN can outperform classical counterpart with close number of trainable parameters \cite{zeng2022multi, QCNN_classical_data}. However, it is not yet known whether optimizing a parametrized quantum circuit can lead to an algorithm that may outperform any state-of-the-art classical algorithm. It is known that even a single-qubit quantum circuit is enough to solve non-trivial classification tasks \cite{spain_class, spain_reg}, there is potential in using quantum kernel estimation for support vector machines \cite{liu2021rigorous, huang2021power}, and it is supposed that PQC-based models may have advantages in expressivity and generalization \cite{raghu2017expressive, schuld_grad, abbas2021power, du2022efficient, caro2022generalization}. Also, PQC unitarity automatically ensures effective weight normalization \cite{mitarai2018quantum}, which is useful for recurrent models. To date, the idea found a few experimental realizations on various physical platforms, the most prominent being trapped ions and superconducting artificial atoms \cite{zhu2019training, havlivcek2019supervised, huang2021experimental, ren2022experimental, dutta2022single}.


To achieve quantum advantage for a certain ML task is to find a PQC than will train or perform inference faster, or with less resources, or with higher accuracy than it is currently possible using classical computational devices. For classification problems, this is achievable probably only with a classically intractable $\bm{\theta}$-parameterized ($\bm \theta \in \mathbbm R^m$) PQC used to map a feature vector $\mathbf{x}\in \mathbbm R^n$ to a higher-dimensional feature Hilbert space vector $\ket{\Phi(\mathbf{x}, \bm{\theta})}$, or, more generally, a density matrix $\rho(\mathbf{x}, \bm \theta)$ lying in the space of unit-trace Hermitian operators \cite{schuld_1,  havlivcek2019supervised, jerbi2023quantum}. Then the prediction is found by performing single- or multi-qubit measurements upon $\rho(\mathbf{x}, \bm \theta)$, optionally using quantum state tomography (QST) \cite{zeng2022multi}. It can be shown that at that last step a linear classifier is formed for $\rho(\mathbf{x}, \bm\theta)$ \cite{schuld2021supervised}. It follows that the classes should be significantly easier to separate in the new feature space than in the original, in similarity to classical dimensionality reduction approaches \cite{van2008visualizing, mcinnes2018umap}.

Pre-processing of $\mathbf x$ and post-processing of measurement outcomes, for example, by simultaneously-trained classical neural networks, seems to improve model performance. As the model combines both classical and quantum mappings, we use the term \textit{hybrid} deep quantum learning model \cite{li2022image, zeng2022multi}. Finally, there may be variations in how the data is inserted in the PQC. Usually, each value from the input data vector $\mathbf{x}$ is written to the parameter of a single gate in the circuit. However, recent studies show that recording the same parameter multiple times over different gates of PQC -- so called \textit{data re-uploading} -- significantly improves its expressivity \cite{spain_class, jerbi2023quantum}.

In this work, we experimentally train a small hybrid model to solve several problems of supervised learning on well-known datasets -- bit-string parity (\textsc{parity}), diagnosing breast cancer by biopsy results (\textsc{cancer}), discerning wine cultivars by their physical and chemical parameters (\textsc{wines}), and recognizing handwritten digits (\textsc{mnist}) \cite{ucidb}. We find this to be the first report of an experiment employing loading image datasets into a PQC via convolutions and classification of all 10 digits from \textsc{mnist} dataset. As a small-scale prototype of a quantum hardware device used to evaluate the quantum part of the model, we use a linear array of superconducting transmon artificial atoms with nearest-neighbour interactions \cite{Koch, barends2013coherent, huang2021experimental}. We show that four qubits with at most two hundred parameters is enough to reach test accuracies higher than $90$\% for all studied datasets. {As we find experimentally and confirm numerically, the algorithm is sufficiently resistant to imperfections in gate operations and can be implemented with currently available quantum devices without error correction.

\begin{figure*}
  \centering
  \includegraphics[width=0.85\textwidth]{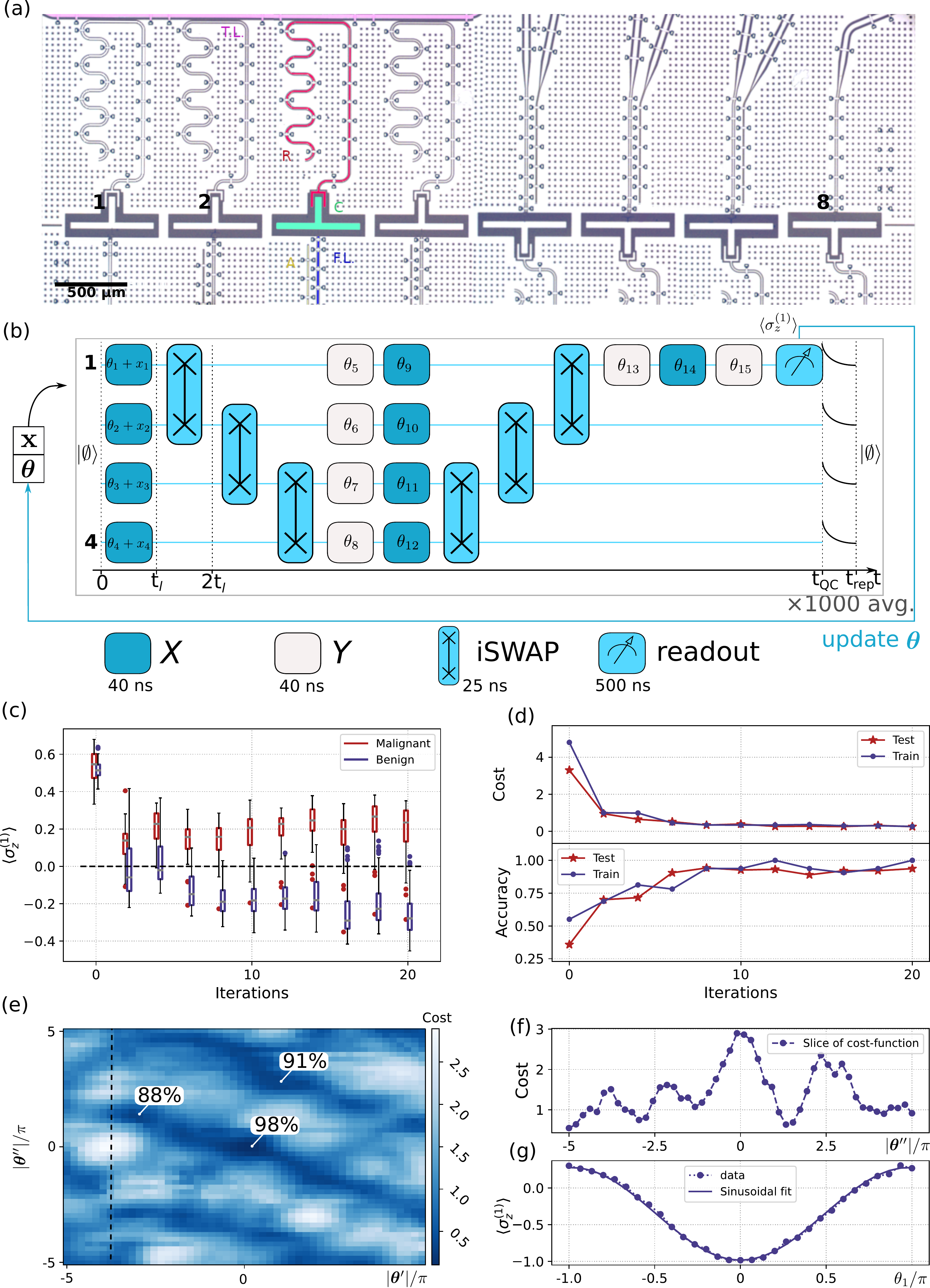}
  \caption{ (a) -- Micrograph of the eight-transmon device used as the quantum hardware (false coloured). The chip is symmetric, one of the two transmission lines (purple) is visible. Readout resonators (red), microwave antennas (yellow) and flux control lines (blue) address each transmon individually. T-shaped shunting capacitors are shown in green. (b) -- Structure of the PQC. The polar angles of single-qubit $X$, $Y$ rotations constitute the parameter vector $\bm \theta$ while two-qubit operations are fixed.  (c) -- Distributions of $\langle\sigma_z^{(1)}\rangle$ for two classes vs. training iteration for \textsc{cancer} dataset. (d) -- Cost and accuracy convergence, calculated for both $\mathcal T$ and $\tilde{\mathcal T}$ for comparison. (e) -- Measured cost function landscape for $\tilde{\mathcal T}$ around the found minimum in the linear hull of two random orthogonal directions $\bm \theta',\ \bm\theta''$. Accuracies are indicated at several local minima. (f) -- A 1D slice of cost function, shown in (e) with a dashed line. (g) -- Output of the circuit $\langle\sigma_z^{(1)}\rangle$ showing expected harmonic dependence on $\bm{\theta}$-components, data for $\theta_1$.}
  \label{figure1}
\end{figure*}

\section{Learning simple datasets}

An optical image of the experimental device is shown in Fig. \ref{figure1}(a). The chip hosts  eight artificial atoms forming a linear chain with nearest-neighbour interactions \cite{huang2021experimental}. Each transmon has an individual measurement resonator, and two control lines -- to change the flux through the SQUID of transmon and to excite it with microwave radiation. In this study, we use only the left half of the chain (4 transmons). The physical parameters of the device are presented in Table \ref{tab_device}.

\begin{table}[b]
\setlength{\tabcolsep}{0pt}
\centering
\begin{tabular}{ p{3cm}   p{0.9cm}  p{0.5cm}  p{0.9cm}  p{0.5cm}  p{0.9cm} p{0.5cm} p{0.9cm}} 
 \hline
Transmon  & 1 && 2 && 3 && 4\\
 \hline
 ${\omega_{ge}}/{2\pi}$, GHz 
 & 5.80 && 4.96 && 6.05 && 4.94\\
 $T_1$, $\mu$s  & $7.2$ && $9.3$ 
 && $8.0$ && $10.5$\\
 $T_{2}^{*}$, $\mu$s 
 & $3.8$ && $4.1$ && $3.7$ && $3.9$\\
 ${\omega_{r}}/{2\pi}$, GHz 
 & 7.16  && 7.24 && 7.31 && 7.40  \\
$t_{X,Y}$, ns & 40 && 40&& 40 && 40\\
 $t_\text{iSWAP}$, ns && 25 && 27 && 27 & \\
 \hline
\end{tabular}
\caption{Physical parameters for the four leftmost transmons. Idling (``parking'') point frequencies $\omega_{ge}^{(i)}$ and readout resonator frequencies $\omega_r^{(i)}$ are given, decoherence times are measured when the transmons are in their idling points, near the ``sweet-spots''. Precise gate durations are also listed.}
\label{tab_device}
\end{table}

\subsection{Training the device}

The architecture of the four-qubit PQC that is run on the device is shown in Figure \ref{figure1}(b). Gaussian driving pulses with controllable amplitude are used for single-qubit operations ($t_{X,Y} = 40$ ns) and smoothed rectangular DC pulses for two-qubit operations ($t_\text{iSWAP}=25$ ns), details of the calibration are presented in Appendix \ref{Appendix1}. Each layer of  operations has a duration of $t_l = 80$ ns, which includes an idling margin to account for variations of propagation delay among the control lines. Total PQC execution time is $t_\text{PQC} = 1460$ ns (incl. 500 ns for readout); however, we have to use a $t_\text{rep} = 50$ $\mu$s repetition period to allow the system to return to the state of thermal equilibrium with the environment of 20 mK, denoted as $\ket{\emptyset}$. This repetition period can be significantly reduced, though, to 1-5 $\mu$s, by using unconditional reset protocols \cite{Reset2013, Reset2}. 

Controlled evolution of a real quantum system ends in a statistically mixed state $\rho(\mathbf x, \bm \theta)$, having limited resemblance to the PQC target state $\ket{\Phi(\mathbf{x}, \bm \theta)}$, and this is supposed to be the main restriction for VQA development on larger NISQ devices \cite{wang2021noise}. However, while in our case the PQC execution time $t_\text{PQC}$ is comparable to the average decoherence time $T_2 \approx 4\ \mu$s, we are still able to successfully perform training and inference.

The feature vectors $\mathbf{x} \in \mathbbm R^4$ are presented to the PQC by single-qubit $X$-rotations of the first layer. The corresponding angles are calculated by applying the inverse tangent function to $\{x_i\}$. For datasets \textsc{cancer} and \textsc{wines}, we use decision tree classifier to choose} four most relevant components of the feature vector. Selected features for \textsc{wines} dataset are ``proline'', ``flavanoids'', ``color intensity'' and ``alcohol''; for \textsc{cancer} dataset -- ``worst radius'', ``worst concave points'', ``worst texture'' and ``mean texture''.  For a certain dataset, we will denote as $\mathcal X$, $\mathcal Y$ the sets of feature vectors and corresponding labels, and as $\mathcal T$, $\tilde{\mathcal T}: \mathcal T \cup \tilde{\mathcal T } = \mathcal X$ the train and test feature subsets, respectively. The information about used datasets is summarized in Table \ref{tab_datasets}.

Additionally, we merge into the first layer four components $\theta_{1-4}$ of the weight vector $\bm \theta \in \mathbbm R^{15}$ by adding them to the respective feature angles \cite{ren2022experimental} This operation is necessary to shorten PQC by combining layers of encoding and optimization.

The larger part of Hilbert space can be reached by the image $\ket{\Phi(\mathbf{x}, \bm{\theta})}$, the better expressivity of the model, and thus the PQC must be sufficient to generate fully entangled states. However, higher expressivity might lead to training problems of the algorithm, so the model should not be too complex. While it is possible to optimize the structure of the circuit along with tuning its parameters $\bm{\theta}$ \cite{du2022quantum}, we use a V-shaped sequence of fixed two-qubit operations, interleaved by two layers of single-qubit $X$ and $Y$ rotations, as the entangling block. This structure allows us to effectively use a large part of the Hilbert space of 4 qubits, has sufficient flexibility and is easy to calibrate. We use only roughly-calibrated quasi-iSWAP gates \cite{dewes2012demonstrating, foxen2020demonstrating}. Their exact matrix representation does not significantly affect the expressivity of the circuit, which we check in numerical simulations and connect with the Kraus-Cirac decomposition theorem \cite{kraus2001optimal}. The detailed information about calibration of single-qubit and two-qubit operations can be found in Appendix \ref{Appendix1}. Fortunately, there is no need to know exactly which final state $\rho(\mathbf{x}_i, \bm{\theta})$ is prepared, contrary to quantum chemistry problems \cite{cerezo2021variational}. As there are many possible structures for PQCs, numerical performance analysis some of them is provided in Appendix \ref{Appendix2}.

The last four layers are an arbitrary Euler rotation of the first qubit via an $Y$-$X$-$Y$ sequence, and a measurement. The prediction $g(\mathbf{x}_i, \bm \theta) = \text{Tr}\left[\rho(\mathbf{x}_i, \bm{\theta}) \sigma_z^{(1)}\right]$ is calculated by running the PQC repeatedly and averaging the $\sigma_z^{(1)}$ outcomes. Then it is thresholded at $g(\mathbf{x}_i, \bm \theta)=0$ to obtain the binary prediction. With such an output, $k$-class classification is possible to realize by training $k$ one-versus-others models with proper optimal parameters $\{\bm \theta_1, ..., \bm \theta_k\}$, or $k(k-1)/2$ pairwise classifiers \cite{li2022image}. 

 We optimize $\bm \theta$ using stochastic gradient descent (SGD) with the per-sample logarithmic cost function defined by 
\begin{equation}\label{eq1}
\begin{aligned}
\mathcal{L}\left[ g(\mathbf{x}_i, \bm \theta), y_i \right]  = &\log_2\left( 1+\exp[ -y_i\cdot g(\mathbf{x}_i, \bm \theta) \cdot \beta]\right) \\
 & + \gamma\cdot|\bm \theta|^2, 
\end{aligned}
\end{equation}
which favours the label $y_i \in \{-1,1\}$ and the prediction $g(\mathbf{x}_i, \bm \theta)$ to have the same sign, and grows linearly in $|g(\mathbf{x}_i, \bm \theta)|$ when they have opposite signs. We find that choosing large $\beta = 10$ to strongly penalize sign difference is beneficial to the training process. The regularizing term $\gamma\cdot|\bm \theta|^2$ with $\gamma = 0.2$ allows to penalize the model for overfitting. While important for the image recognition task below, it is not necessary for simple datasets. We also check that for the studied datasets, the quadratic cost yields comparable training performance.

The full cost is calculated as the expectation value $\mathbbm E \left[\mathcal{L}\left[ g(\mathbf{x}_i, \bm \theta), y_i \right]\right]$ over a chosen data subset $\mathcal S$, so $i: \mathbf{x}_i \in \mathcal S$, with $\mathcal S$ being $\mathcal T$, or $\tilde{\mathcal T}$, or a mini-batch $\mathcal B$ of size $b$. The $j$-th component of its gradient over $\bm \theta$ can be conveniently computed using the parameter-shift rule \cite{mitarai2018quantum, schuld_grad} which requires only two measurements at $\theta_j \pm \pi/2$. We use the Pennylane library \cite{bergholm2018pennylane} with a custom software wrapper to our experimental setup to perform both the automated differentiation and the optimization.

\begin{table}
\centering
\begin{tabular}{lcccc} 
 \hline
 Dataset & \textsc{parity} & \textsc{cancer} & \textsc{wines} & \textsc{mnist}\\
 \hline
 \# samples & 16 & 569 & 178 & 1797 \\
 \# features & 4 & 30 & 13 & 56 \\
 \# classes & 2 & 2 & 3 & 10\\
  Accuracy & $1.0$ & $0.95^*$ & $0.94^*$ & $0.90$\\
 \hline
\end{tabular}
\caption{Summary of the dataset properties. For \textsc{parity}, due to the low number of samples for a 4-bit task train/test split was equal. For the remaining two datasets the splitting was 2/1. $^*$ cross-validated accuracy, averaged over 6 different random splits}
\label{tab_datasets}
\end{table}

\begin{figure*}
  \centering
  \includegraphics[width=1.0\textwidth]{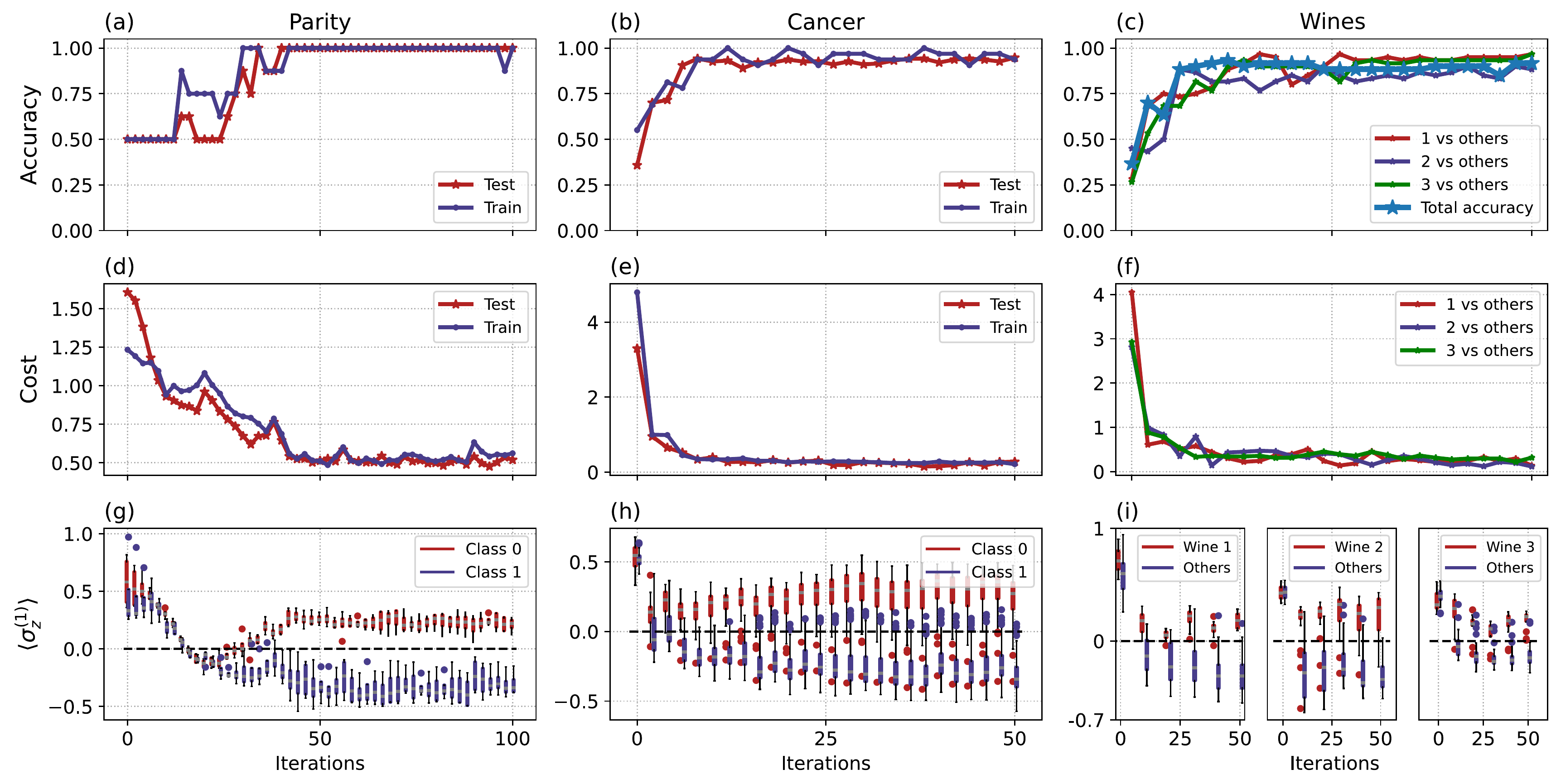}
  \caption{Training process of classifiers for all three simple problems. (a)-(c) -- Accuracy vs. the number of iterations. (d)-(f) -- Same for the cost function. (g)-(i) --  Distributions of $\langle\sigma_z^{(1)}\rangle$ for two classes during training process. For wines dataset accuracies of all 3 ``one-vs-others'' classifiers and total accuracy are plotted. In picture (i) the distribution of $\langle\sigma_z^{(1)}\rangle$ for these classifiers is plotted.}
  \label{figure2}
\end{figure*}

Figure \ref{figure1}(c, d) shows a visualization of the training process for the \textsc{cancer} dataset with 569 samples and 2-to-1 $\mathcal{T}$-to-$\tilde{\mathcal T}$ split \cite{dobbin2011optimally}. Figure \ref{figure1}(c) shows the distribution of the model predictions for $\mathbf x \in \tilde{\mathcal T}$ at each training step. Each iteration consists of one gradient evaluation and a Nesterov accelerated \cite{nesterov1983method} SGD step over $x_i \in \mathcal B$, $b = 64$. At the beginning of the training, the two classes are indistinguishable while at the end of the training algorithm the distributions of $g(\mathbf x, \bm \theta)$ for the two classes almost do not overlap. The accuracy of the algorithm on both $\mathcal{T}$ and $\tilde{\mathcal T}$ steadily increases with the number of iterations, reaching approx. 95\% in about 10 iterations. We also do not observe any systematic decline in accuracy if the training is further continued. To correct for the statistical fluctuations of the accuracy estimation due to the particular realization of the sampling for the $\mathcal{T}$-to-$\tilde{\mathcal T}$ split, we use the cross-validation method, averaging results over several different splits.

Following \cite{anschuetz2022quantum}, we also study the behaviour of the cost function  calculated over $\mathbf x \in \tilde{\mathcal T}$ near the found optimum. Using a known visualization method \cite{li2018visualizing}, we plot a 2D slice of the 15D parameter space, a square in the linear hull of two normalized orthogonal random vectors $\bm \theta',\ \bm \theta''$ added to the optimal vector $\bm \theta^*$. From Figure \ref{figure1}(e) it can be seen that even for such a small PQC, the minimum indeed is not unique which can lead to trapping of the algorithm, and that the cost is non-convex in the original space \cite{li2018visualizing}. We also find that setting $\bm \theta^* = 0$ in this experiment yields similar topography with local minima corresponding to above 80\% accuracies, so a moderately good solution can be found just by moving along a randomly chosen direction.

We also check experimentally that the dependence of $g(\mathbf{x}_i, \bm \theta)$ on each of the parameters $\theta_j$ is harmonic, according to theory, and show in Figure \ref{figure1}(g) how $g(\mathbf{x}_i, \bm \theta)$ varies with $\theta_1$, the first and the deepest parameter in the circuit, when the other parameters and the input features $\mathbf x$ are set to zero. As can be seen, due to the inaccuracies in the calibration of the two-qubit gates and non-negligible decoherence, the value of the prediction never reaches ``+1'' in contrast to what one would expect from the PQC structure for $\theta_1 = \pi$.

\subsection{Performance analysis}

In Figure \ref{figure2} we summarize the model training and performance on all three datasets, see also Table \ref{tab_datasets}. In the \textsc{parity} problem, 4-bit sequences should be decided to contain even or odd number of ``1''-s. This problem is a simple test which displays the reproducibility of quantum operations and the sensitivity of the model in capturing class change even when a single bit is flipped, which is difficult for classical models \cite{de2019random}. Having only 16 samples, we split the dataset equally and calculate the cost function on the full subset $\mathcal T \equiv \mathcal B$, $b=8$. At the optimal point, the accuracy reaches 100\%, as there exists an analytical solution composed of four CNOT gates \cite{riste2017demonstration} which can be mimicked by our ansatz. At the same time, the cost function does not reach zero due to the imperfections of quantum operations and decoherence.

The \textsc{cancer} problem, already briefly presented above, is also binary, but has a significantly larger dataset which allows a better splitting ratio and puts the model under a more stringent test in terms of performance. As it can be seen from Figure \ref{figure1}(b), with $t_\text{rep} = 50$ $\mu$s and averaging over 1000 repetitions, the measurement of $\langle \sigma_{z}^{(1)} \rangle$ takes 50 ms time. In our setup, rewriting the pulse sequence waveforms to update the gates in the PQC takes comparable time (it could be reduced at least an order of magnitude, though, with better hardware). To find the gradient of the cost function calculated on a single $x_i$, it is necessary to measure $\langle \sigma_{z}^{(1)} \rangle$ for $2m + 1$ sets of angles $\bm \theta \in \mathbbm{R}^m$. Evaluating the gradient for the circuit from Figure \ref{figure1}(b) over $m$=15 variables takes $t_\text{grad}=1.55$ s, which with the additional time for rewriting the controlling sequences gives about 3 s. Then, for a batch size $b=64$, one iteration takes around $b\cdot t_\text{grad}=$ 3 min. Finally, reaching the accuracy plateau in around 20 iterations, as in Figure \ref{figure1} (c,d), takes around 1 h.

As an elementary test of the model capability to solve multilabel classification problems, we use the three-class \textsc{wines} dataset. Multilabel classification is done by training three ``one-vs-others'' binary classifiers aiming to detect each of the cultivars. Then, for a given $x \in \tilde{\mathcal T}$, we choose among the three found $\bm\theta_{1-3}$ the one delivering the highest value to $g(\mathbf{x}, \bullet)$, and predict the class accordingly. We find that all of three one-vs-others classifiers exhibit similar training behaviour - an accuracy of classification starts from approximately 1/3, reaches value of 90\% in 10 iterations and slightly fluctuates further, which is normal for mini-batch learning. As a result, the total classification accuracy is also slightly above 90\%; the cross-validation procedure gives a value of 94\%.

\begin{figure*}[t]
  \begin{center}
  \centering
  \includegraphics[width=.9\textwidth]{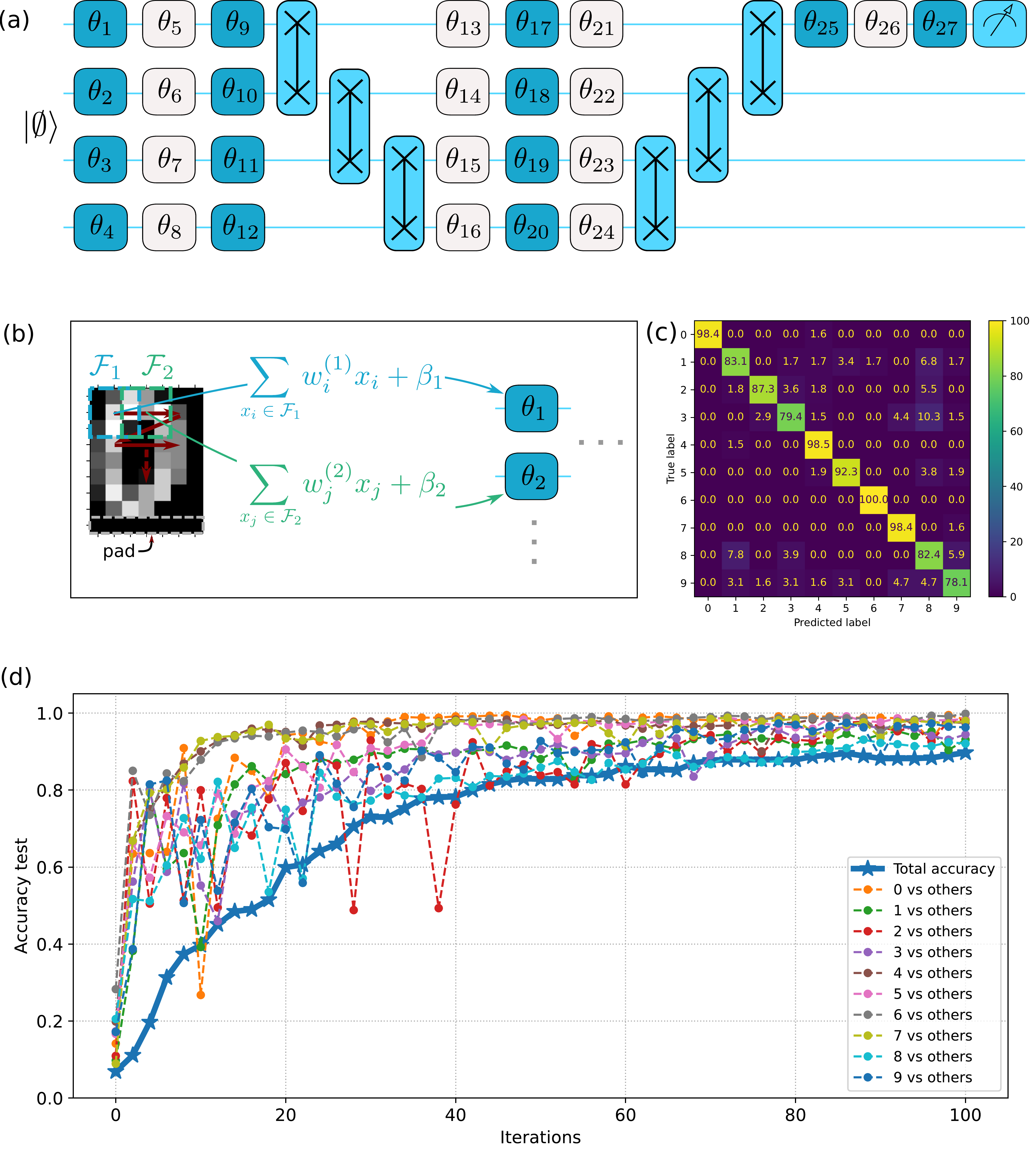}
  \end{center}
  \caption{Image recognition for \textsc{mnist}. (a) -- The PQC used to process larger feature vectors. (b) -- Data encoding into parameters of single-qubit operations, using convolutional kernels. (c) -- Confusion matrix to analyze the performance of each classifier. Each intersection of $i$-th row and $j$-th column shows the percentage of pictures belonging to the $i$-th class and recognized as elements of $j$-th class. Non-diagonal elements show misclassifications. (d) -- Visualization of training process of the classifier. The dependence of the accuracies on the number of iterations for 10 one-vs-others classifiers and total accuracy are shown.}
  \label{mnist}
\end{figure*}

\section{Recognition of handwritten digits} \label{image recognition}
The feature space dimension $m=4$ for the datasets considered above is obviously quite small. The situation is different for the image recognition problem: even for a downsampled and cropped \textsc{mnist} picture of size $8\times 7$ pixels with intensities ranging from 0 to 1, feature space is equipotent to $\mathbbm R^{56}$.  Choosing a particular way to load information from this space into the quantum state of just four qubits is not a trivial task. We use an approach combining the data re-uploading concept \cite{spain_class} and convolutional neural networks (CNN) \cite{lecun1998gradient, zeng2022multi}.   For pre-processing we use normalisation and inverse tangent transformation of the features. We use a modified PQC with the structure similar to the one shown in Figure \ref{figure1}(a), given in Figure \ref{mnist}(a). In Figure \ref{mnist}(b), we illustrate how the $8\times 7$ images are padded with a zero bottom row and then divided in $3\times 3$ partially overlapping local receptive fields (LRF) moving with stride $2$ in both directions \cite{lecun1998gradient}. LRFs traverse the image twice, so there are 24 weight kernels in total. Pixels belonging to the $i$-th LRF $\mathcal F_i$ are convolved with weights $w^{(i)}_j$, and, with addition of a bias $\beta_i$, are converted to angles $\theta_{1-24}$:
\[
\theta_i = \beta_i + \sum_{x_j \in \mathcal F_i} w^{(i)}_j x_j .
\] 
 In contrast to conventional CNN, the kernels for each LRF are independent, which makes the model more flexible. The last three parameters $\theta_{25-27}$ are independent, so with an addition of a final bias $\beta_0$ to the output of the PQC, the total weight dimension is $244$. We use parameter-shift rule to compute $\bm\theta$-gradients of the loss-function and chain rule to obtain $w^{(i)}_j$ and $\beta_i$ gradients. 

In contrast to the circuit in Figure \ref{figure1}(a), the feature data are now recorded in multiple layers. The information about every pixel is also written several times. 

The model was tested on a subset of the \textsc{mnist} dataset, see Table \ref{tab_datasets}, by training 10 separate one-vs-others classifiers to be able to distinguish between all 10 digits. There is a strong (1/9) disproportionality in the quantities for each of the corresponding pairs of classifiers, which might lead to training problems. To overcome these problems, we perform dataset balancing by adding copies of elements of a smaller class with added Gaussian noise to the training dataset. After that the number of elements in the two classes becomes the same. The training process is visualized in Figure \ref{mnist}(d): the total 10-digit accuracy steadily increases from approximately 10\% in the beginning (random guessing) to approx. 90\% after 100 iterations. The accuracy of individual classifiers varies from 100\% for `6 vs. others' to 92\% for `8 vs. others', which decreases the full accuracy to 90\%. To analyze the errors, we construct a confusion matrix, which is shown in Figure \ref{mnist}(c), and find that most confusion is caused by the ``8'' classifier, with most misclassifications between ``8'' and ``3'', which looks reasonable. Probably, the accuracy could be further improved, however as the training for Figure \ref{mnist} had taken around 100 h., we did not continue it further. We found in numerical simulations that suggested PQC is not optimal, and classification accuracy might be increased to approximately 93\% with nearly the same number of circuit layers. The detailed analysis of possible circuit architectures is presented in Appendix \ref{Appendix2}.
We have also tested the model on the \textsc{fashion mnist} dataset \cite{fashion_mnist}. As for the classical models \cite{fashion_mnist}, the classification accuracy for the \textsc{fashion mnist} dataset turned out to be worse than for the \textsc{mnist} dataset. We achieve only 85\% accuracy for 4 different types of clothes, while for four digits ``0 - 3'' we report 98\% accuracy.

\section{Conclusion}
We experimentally implement a supervised quantum learning algorithm in a chain of superconducting qubits to solve multilabel classification and image recognition problems. We firstly realize loading image datasets onto PQC via convolutions, present a suitable gate sequence and a training algorithm. This algorithm allows us to achieve classification accuracy $90\%$ for the \textsc{mnist} dataset. Our classification algorithm is convenient for implementation on near-term quantum devices without error correction because it is sufficiently resistant to imperfections and requires measurement of only one qubit. We note that the presented model does not yet outperform even the simplest classical model, such as linear classifier, which achieves an accuracy of 95\% with only 570 trainable parameters. This means, though, that it is possible to obtain 95\% accuracy using a PQC with only 1 layer and 1 qubit by training 10 one-vs-others classifiers if a linear combination of all features is recorded in the angle of single-qubit operations. However, the main work in that case will be performed by a classical computer, while to achieve advantage in quantum machine learning the right balance between the classical and quantum parts of the model should be found. For example, replacing the last low-dimensional layers of a convolutional network with a PQC could be a direction of further study. 

We also address the issue of the low speed of inference that we observe in practice with a real device. Despite the fact that we use state-of-the-art gate durations (10-s of ns), and the fact that the superconducting quantum computing platform currently features the fastest known gates (compared with silicon, with 100-s of ns \cite{philips2022universal}, and trapped atom/ion or diamond platforms, with 100-s of $\mu$s \cite{postler2022demonstration, bluvstein2022quantum, abobeih2022fault}), the training process is currently orders of magnitude slower than for the classical machine learning methods. We note, however, that the training time of the hybrid model presented here could be significantly reduced by implementing an unconditional reset instead of simple waiting \cite{Reset2013, Reset2} (about 50 times faster) and training only one multilabel classifier using multiplexed readout instead of 10 binary classifiers. Thus, the training of our model for the \textsc{mnist} dataset could be reduced from 100 hours to approx. 12 min., which is more competitive. It is also possible to reduce total number of layers in the PQC by performing several two-qubit operations in parallel. 

To solve more complex tasks in the domain of quantum machine learning it is necessary to realize very fast gates on a supremacy-scale register of qubits. While in this work we do not notice significant impact of gradient decay on the performance of quantum classifiers, increasing the number of qubits will require an ingenious circuit architecture to cope with that problem. In the classical machine learning, a similar problem has been overcome by using skip connections \cite{he2016deep} and batch normalization \cite{ioffe2015batch}, but at the moment it is not known whether any analogs of these techniques could be reasonably implemented in the quantum case, and further research is necessary. 

\section {Acknowledgments}
The work was partially supported by contract no. RQC-2 dated July 14, 2022 between MIPT and RQC, and by Russian Science Foundation (project no. 21-72-30026). Authors thank E. Korostylev, N. Abramov, A. Strelnikov for valuable technical support, and the anonymous referees for the valuable comments. All samples were fabricated in the Shared Facility Center of MIPT. All authors declare no competing interests. Experimental data are available upon reasonable request from the corresponding authors.

\appendix

\section {Single-qubit and two-qubit operations}
\label{Appendix1}
During the training process, we use parameterized single-qubit operations and fixed two-qubit operations. Single-qubit gates with variational parameters were implemented by using fixed duration (40 ns) microwave pulses with Gaussian envelope. To change the rotation angle $\theta$, we vary the amplitude of pulses, as $\theta$ is proportional to the area of their envelope. To change the rotation axis (X or Y) we vary the phase of pulses (choosing 0 or $\pi / 2$, respectively). 

\begin{figure*}[b]
  \begin{center}
  \centering
  \includegraphics[width=.9\textwidth]{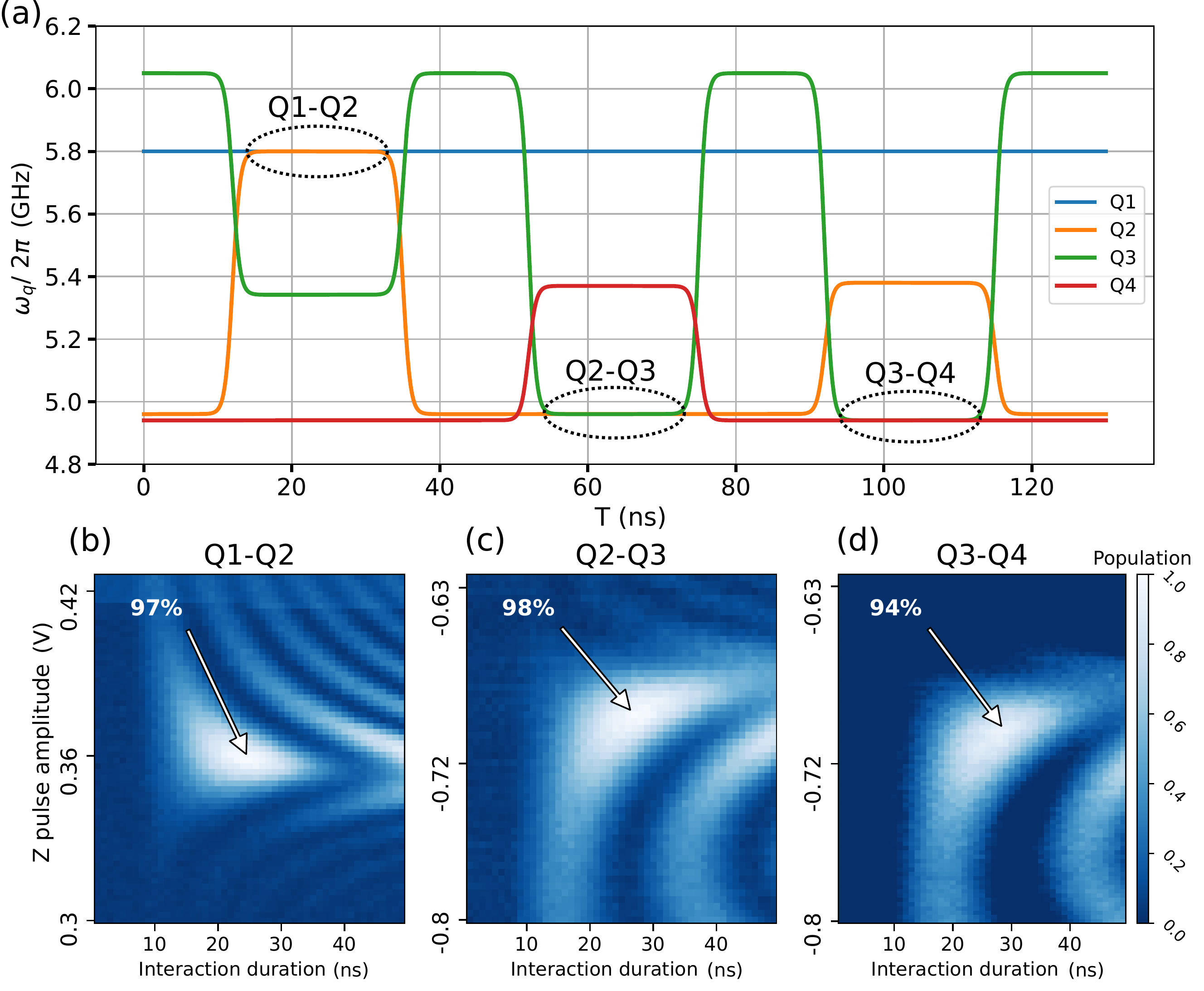}
  \end{center}
  \caption{Calibration of two-qubit operations. \textbf{(a)} Frequency trajectories of the four transmons during three sequential i-Swap operations for neighbouring pairs. When frequencies of one pair are brought into resonance, frequencies of the adjacent transmons are shifted to reduce residual interaction. The areas in which resonances occur are highlighted with dotted ellipses. \textbf{(b)}-\textbf{(d)} Calibration of i-Swap gates. For pictures \textbf{(b)}, \textbf{(c)}, \textbf{(d)} excited qubits are Q1, Q2, Q3 and measured qubits are Q2, Q3, Q4, correspondingly. The dependence of the first excited state population of measured qubit on interaction duration and flux pulse amplitude is shown. The points with the largest population exchange are marked, and corresponding parameters are used for two-qubit operations.}
  \label{2Q_ops}
\end{figure*}

For realization of two-qubit operations, the transmon frequencies are tuned into resonance by fast flux pulses, as shown in Figure \ref{2Q_ops} (a); the resonance regions are highlighted with ellipses. Taking into account the capacitive interaction between the transmons, this procedure leads to the realization of a so-called ``f-Sim''-like two-qubit gate \cite{foxen2020demonstrating}. It consists of single-qubit rotations due to changes in the frequency of the transmons, and of a two-qubit part. As we include single-qubit rotations before and after two-qubit gates in the PQC, the single-qubit part can be automatically compensated during training process. Thus we can limit ourselves to considering only the two-qubit part of the gate: 

\begin{equation}
\text{f-Sim}(\theta,\ \phi)= \begin{pmatrix}
1 & 0 & 0 & 0\\
0 & \cos \theta & -i\sin \theta & 0 \\
0 & -i \sin \theta & \cos \theta & 0\\
0 & 0 & 0 & e^{-i \phi} 
\end{pmatrix}.
\end{equation}  

To calibrate the flux pulses, we measure two-qubit chevron-type oscillations. In this experiment, one of the transmons in the pair is excited by a $\pi$-pulse and then brought close to resonance with its pair transmon. The population of the latter is recorded vs. amplitude and duration of the flux pulse. The resulting dependencies for the three pairs of adjacent transmons are shown in Figure \ref{2Q_ops} (b--d). We maximize the population transfer, so that $\theta \approx \pi/2$. We also directly measure the non-zero phase $\phi$ due to ZZ interaction to be $\approx 0.1\pi$ in a separate Ramsey-type  experiment. 

To address the robustness of the PQC performance to inaccuracies of the parameters $\theta$ and $\phi$, we numerically study the dependence of the accuracy of the classification for the \textsc{cancer} dataset on $\theta$ and $\phi$. This result is shown in Figure \ref{Cancer_circuits}(f). It can be seen that the classification accuracy remains almost unchanged, and the convergence is preserved for $\theta \in [0.2 \pi,\ 0.8 \pi]$ and $\phi \in [-0.5 \pi,\ 0.5 \pi]$. 

\section{Alternative circuit architectures}
\label{Appendix2}
\begin{figure*}[h]
  \begin{center}
  \centering
  \includegraphics[width=.9\textwidth]{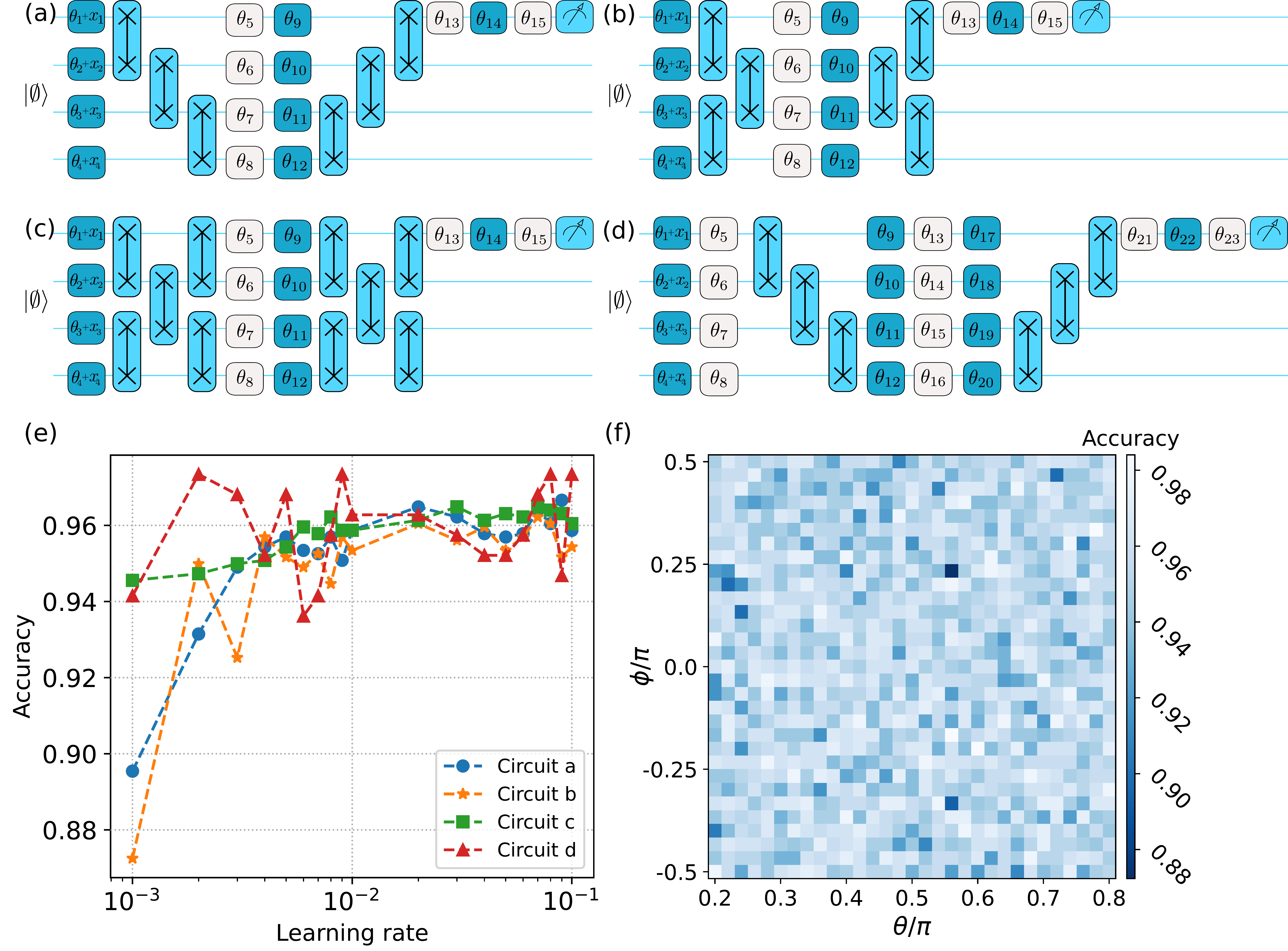}
  \end{center}
  \caption{Comparison of the different PQCs architectures for classification of \textsc{cancer} dataset. (a)-(d) -- Possible structures of PQCs. (e) -- Numerical modelling of the accuracy vs. learning rate dependence for considered PQCs. (f) -- Numerical modelling of the accuracy vs. angles $\theta$ and $\phi$ dependence for PQC (a) and dataset \textsc{cancer}.}
  \label{Cancer_circuits}
\end{figure*}

During our investigation we were focused on only one PQC architecture for simple datasets and one architecture for more complex \textsc{mnist} dataset. The chosen architecture provides a sufficient degree of entanglement using a relatively small number of layers. We can also consider PQC in which two-qubit operations between qubits 1,2 and 3,4 are done simultaneously. Possible structures of these PQCs are shown in Figure \ref{Cancer_circuits} (a--d). We numerically tested the ability of these PQCs to solve binary classification problems using \textsc{cancer} dataset. The dependence of cross-validation accuracy on learning rate for these circuits is shown in Figure \ref{Cancer_circuits}(e). As it can be seen, the best accuracy does not depend significantly on the choice of PQC. For experimental realization it is convenient to use PQC without simultaneous two-qubit operations. This is the reason why we choose PQC (a) for our experiments. 

\begin{figure*}[h]
  \begin{center}
  \centering
  \includegraphics[width=.9\textwidth]{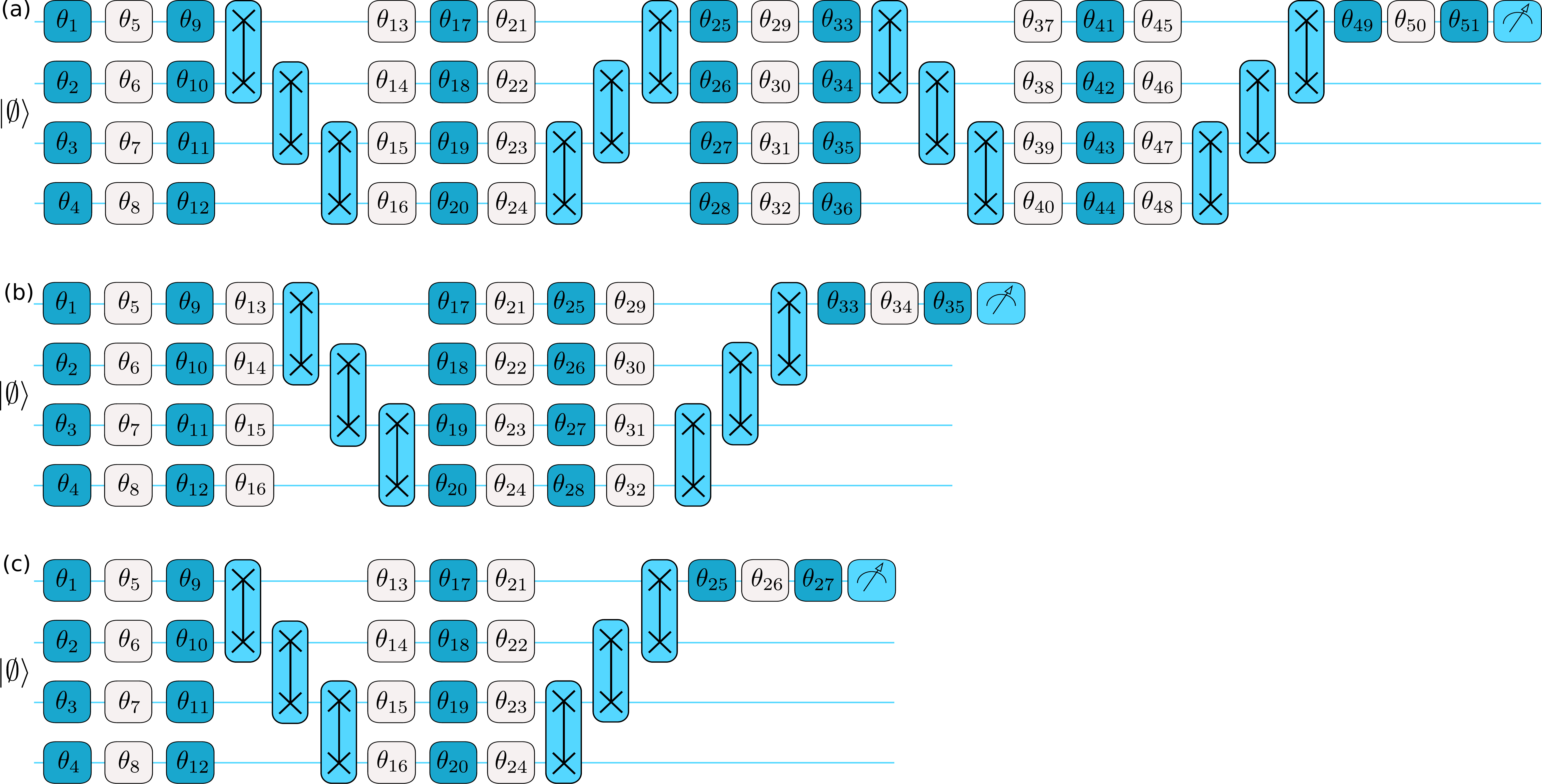}
  \end{center}
  \caption{Possible PQC architectures for the image recognition problem.}
  \label{MNIST_circuits}
\end{figure*}

The PQC for image recognition has many control parameters, for example, size of LRF or stride. These parameters may strongly affect the final classification accuracy.  Possible architectures of PQCs are shown in Figure \ref{MNIST_circuits}. To find the optimal circuit, we conducted numerical simulations for various PQCs and control parameters. Results of our numerical modelling are summarized in Table \ref{MNIST_PQCs_table}. It can be noticed that chosen for experimental realization architecture of PQC (Figure \ref{MNIST_circuits}(c)) with size of LRF 3x3 and stride 2 does not allow to achieve the highest accuracy of classification. Using circuit with nearly the same number of layers, total accuracy might be improved to approximately 93\%.

\begin{table}[h]
\centering
\begin{tabular}{cccccc} 
\hline
PQC & LRF size & stride & parameters &  layers &  accuracy\\
\hline
a & 2x2 & 1 &  248 & 27 & 0.95\\
c & 2x2 & 2 &  92 & 15 & 0.925\\
b & 2x2 & 2 & 164 & 17 & 0.93\\
c & 3x3 & 2 &  244 & 15 & 0.9 \\
a & 3x3 & 1 & 376 & 27 & 0.77 \\
\hline
\end{tabular}
\caption{Comparison of different PQC architectures for image recognition problem. In the first column the corresponding circuit label from Figure \ref{MNIST_circuits} is shown.}
\label{MNIST_PQCs_table}
\end{table}

\bibliography{bibliography}

\end{document}